\documentclass[epj]{webofc}
\usepackage[varg]{txfonts}   
\woctitle{XLV International Symposium on Multiparticle Dynamics}
\begin{document}

\newcommand{\bkt}{\mbox{\boldmath $k_{t}$}}
\newcommand{\dilepton}{\mbox{\boldmath $e^{+}e^{-}$}}
\newcommand{\p}{\partial}
\newcommand{\bsigma}{\mbox{\boldmath $\sigma$}}
\newcommand{\btau}{\mbox{\boldmath $\tau$}}
\newcommand{\brho}{\mbox{\boldmath $\rho$}}
\newcommand{\bpipi}{\mbox{\boldmath $\pi\pi$}} 
\newcommand{\bss}{\bsigma\!\cdot\!\bsigma}
\newcommand{\btt}{\btau\!\cdot\!\btau}
\newcommand{\bnabla}{\mbox{\boldmath $\nabla$}}
\newcommand{\bphi}{\mbox{\boldmath $\tau$}}
\newcommand{\bvarphi}{\mbox{\boldmath $\rho$}}
\newcommand{\bDelta}{\mbox{\boldmath $\Delta$}}
\newcommand{\bpsi}{\mbox{\boldmath $\psi$}}
\newcommand{\bPsi}{\mbox{\boldmath $\Psi$}}
\newcommand{\bPhi}{\mbox{\boldmath $\Phi$}}
\newcommand{\bnab}{\mbox{\boldmath $\nabla$}}
\newcommand{\bpi}{\mbox{\boldmath $\pi$}}
\newcommand{\btheta}{\mbox{\boldmath $\theta$}}
\newcommand{\bkappa}{\mbox{\boldmath $\kappa$}}
\newcommand{\bn}{\mbox{\boldmath $n$}}
\newcommand{\bp}{\mbox{\boldmath $p$}}
\newcommand{\bq}{\mbox{\boldmath $q$}}
\newcommand{\br}{\mbox{\boldmath $r$}}
\newcommand{\bs}{\mbox{\boldmath $s$}}
\newcommand{\bk}{\mbox{\boldmath $k$}}
\newcommand{\bl}{\mbox{\boldmath $l$}}
\newcommand{\bb}{\mbox{\boldmath $b$}}
\newcommand{\bP}{\mbox{\boldmath $P$}}
\newcommand{\bM}{\mbox{\boldmath $M$}}
\newcommand{\ket}[1]{| {#1} \rangle}
\newcommand{\bra}[1]{\langle {#1} |}
\newcommand{\be}{\mbox{\boldmath $e$}}

\newcommand{\rpaeq}{\beq
\left ( \begin{array}{cc}
A&B\\
-B^*&-A^*\end{array}\right )
\left ( \begin{array}{c}
X^{(\kappa})\\Y^{(\kappa)}\end{array}\right )=E_\kappa
\left ( \begin{array}{c}
X^{(\kappa})\\Y^{(\kappa)}\end{array}\right )
\eeq}
\newcommand{\ave}[1]{\langle {#1} \rangle}
\newcommand{\half}{{1\over 2}}
\def\Pom{{\bf I\!P}}
\def\Reg{{\bf I\!R}}

\title{Recent progress in some exclusive and semi-exclusive processes
 in proton-proton collisions}
%
%

\author{Antoni Szczurek\inst{1,2}\fnsep
        \thanks{\email{antoni.szczurek@ifj.edu.pl}}
       \and
        Anna Cisek\inst{2}\fnsep
        \thanks{\email{acisek@univ.rzeszow.pl}}
        \and
        Marta {\L}uszczak\inst{2}\fnsep
        \thanks{\email{luszczak@univ.rzeszow.pl}} 
        \and
        Wolfgang Sch\"afer\inst{1}\fnsep
        \thanks{\email{Wolfgang.Schafer@ifj.edu.pl}}
}

\institute{Institute of Nuclear Physics, Krak\'ow, ul. Radzikowskiego 152
\and
Rzeszòw University, Rzesz\'ow, ul. Rejtana 16}

\abstract{%
We present the main results of our recent analyses of
exclusive production of vector charmonia ($J/\psi$ and $\psi'$)
in $k_t$-factorization approach and
for $\gamma \gamma$ production of charged dilepton pairs in exclusive 
and semiinclusive processes in a new approach, similar in spirit to
$k_t$-factorization. 

The results for charmonia are compared with
recent results of the LHCb collaboration. We include some helicity
flip contributions and quantify the effect of absorption correction.
The effect of $c \bar c$ wave function is illustrated.

We present uncertainties related to $F_2$ structure function which are
the main ingredient of the approach. Our results are compared
with recent CMS data for dilepton production with lepton isolation cuts
imposed.
}
\maketitle
%

\section{Introduction} 

Here we briefly summarize our recent results for exclusive production
of $J/\psi$ mesons obtained in \cite{SS2007,CSS2015} and for
photon-photon production of charged dileptons obtained in
\cite{SFPSS2015,LSS2015}.

\section{Exclusive production of vector charmonia}

\begin{figure}
\begin{center}
\includegraphics[height=4.0cm]{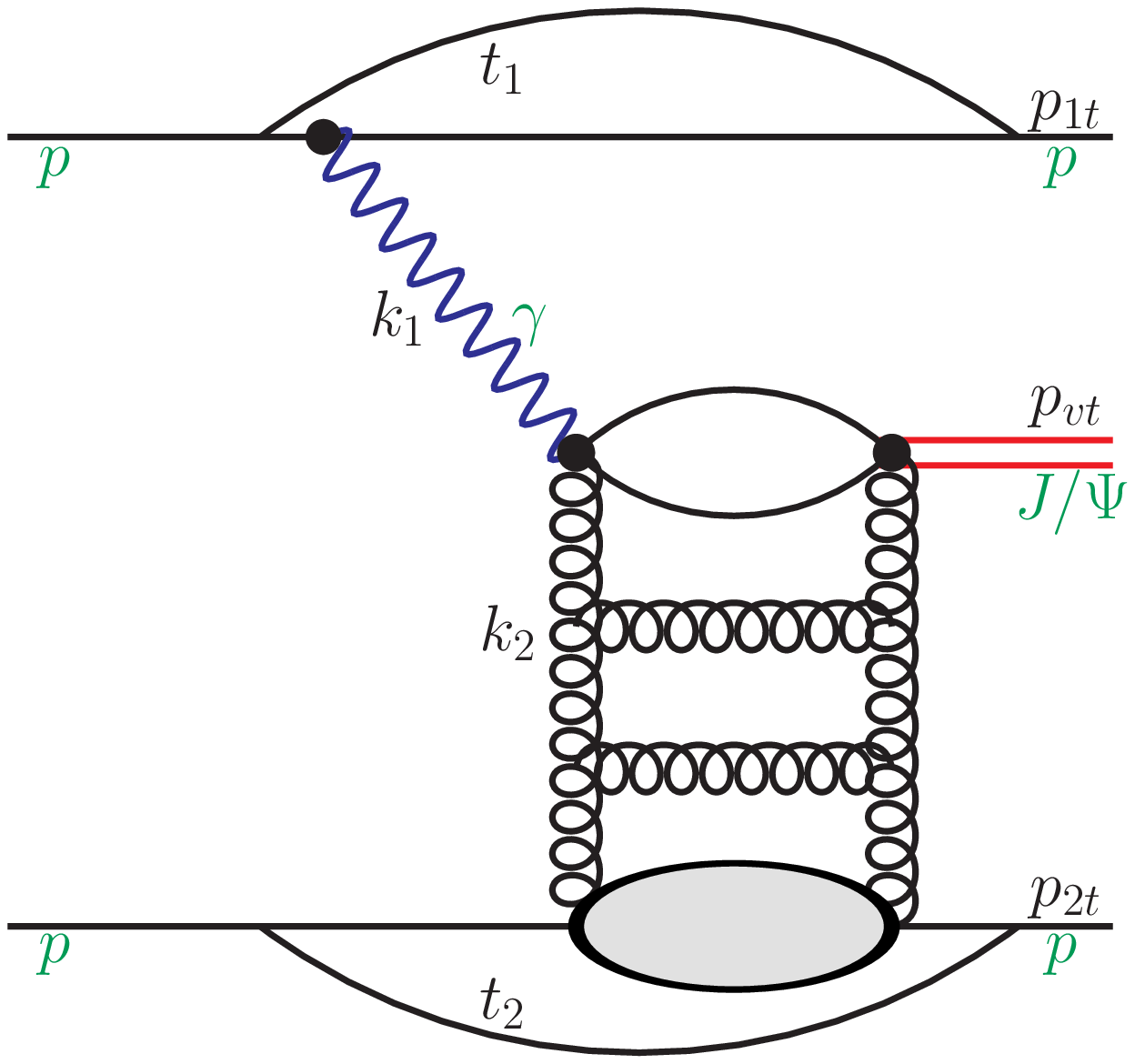}
\includegraphics[height=4.0cm]{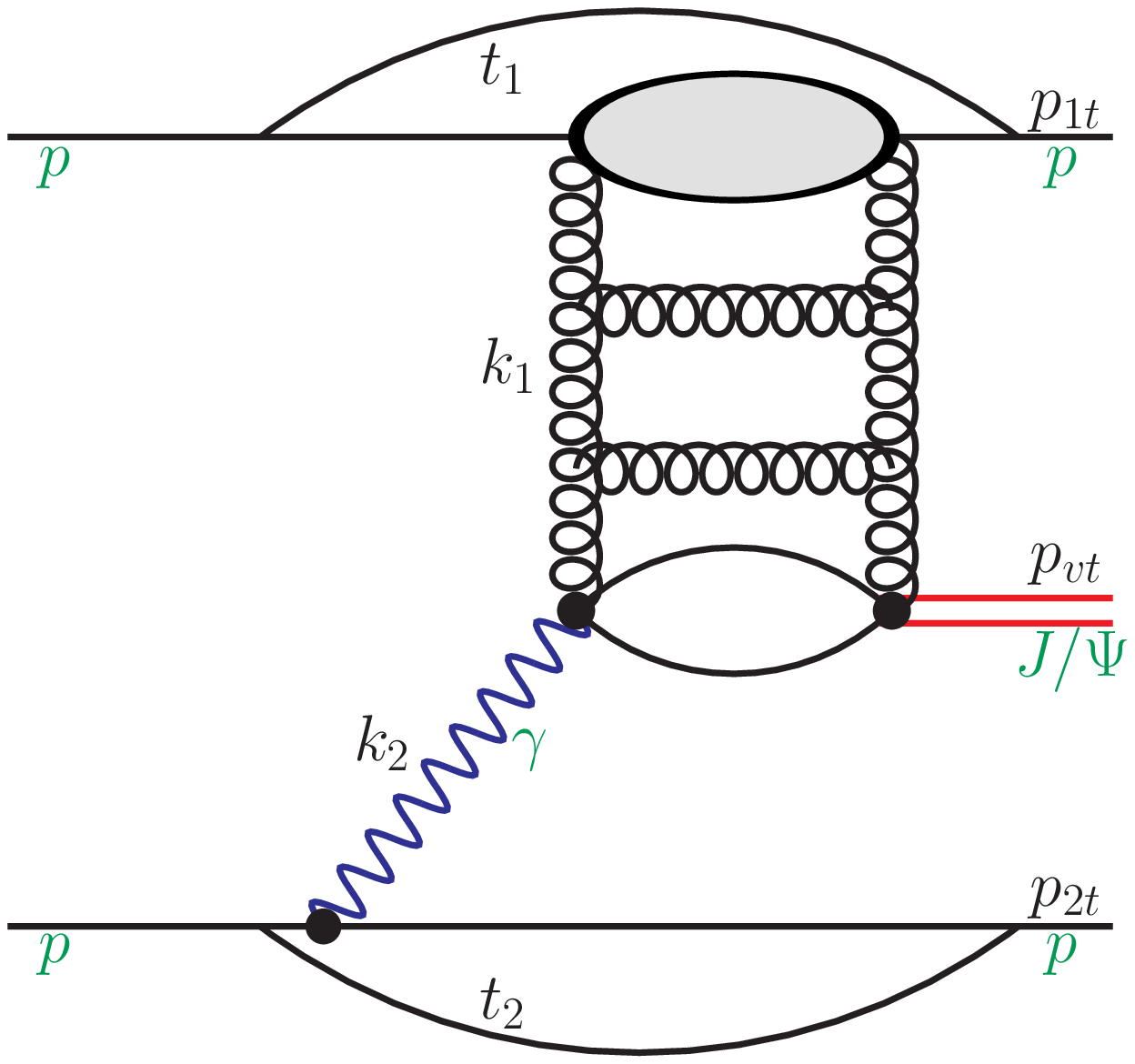}
\end{center}
\caption{Born diagrams for exclusive production of $J/\psi$ mesons.}
\label{fig:diagrams_exclusive_Jpsi}
\end{figure}

The Born diagrams for the exclusive production of $J/\psi$ mesons
are shown in Fig.\ref{fig:diagrams_exclusive_Jpsi}.
In actual calculations in \cite{SS2007,CSS2015} we include also
absorption effects due to soft proton-proton interactions.

All details of the formalism can be found in Refs.\cite{SS2007}
and \cite{CSS2015}. Here we only sketch the main points.

Imaginary part of the forward $\gamma p \to J/\psi p$ amplitude 

\begin{eqnarray}
\Im m \, {\cal M}_{T}(W,\Delta^2 = 0,Q^{2}=0) =
W^2 \frac{c_v \sqrt{4 \pi \alpha_{em}}}{4 \pi^2} \, 2 \, 
  \int_0^1 \frac{dz}{z(1-z)}
 \int_0^\infty \pi dk^2 \psi_V(z,k^2)
 \nonumber \\
 \int_0^\infty
  {\pi d\kappa^2 \over \kappa^4} \alpha_S(q^2) 
 {\cal{F}}(x_{\rm eff},\kappa^2)
 \Big( A_0(z,k^2) \; W_0(k^2,\kappa^2) 
      + A_1(z,k^2) \; W_1(k^2,\kappa^2)
 \Big) \, .
\end{eqnarray}

The full amplitude, at finite momentum transfer is parametrized as:
\begin{eqnarray}
{\cal M}(W,\Delta^2) = (i + \rho) \, \Im m {\cal M}(W,\Delta^2=0,Q^{2}=0)
\, \exp(-B(W) \Delta^2/2) \, ,
\label{full_amp}
\end{eqnarray}
Then the amplitude for the $p p \to p p J/\psi$ can be written somewhat
formally as:
\begin{eqnarray}
{\cal M}_{h_1 h_2 \to h_1 h_2 V}^
{\lambda_1 \lambda_2 \to \lambda'_1 \lambda'_2
  \lambda_V}(s,s_1,s_2,t_1,t_2) =
{\cal M}_{\gamma \Pom} + {\cal M}_{\Pom \gamma} \nonumber \\
= \bra{p_1', \lambda_1'} J_\mu \ket{p_1, \lambda_1} 
\epsilon_{\mu}^*(q_1,\lambda_V) 
{\sqrt{ 4 \pi \alpha_{em}} \over t_1}
{\cal M}_{\gamma^* h_2 \to V h_2}^{\lambda_{\gamma^*} \lambda_2 \to \lambda_V \lambda_2}
(s_2,t_2,Q_1^2)   \nonumber \\
 + \bra{p_2', \lambda_2'} J_\mu \ket{p_2, \lambda_2} 
\epsilon_{\mu}^*(q_2,\lambda_V)  
{\sqrt{ 4 \pi \alpha_{em}} \over t_2}
{\cal M}_{\gamma^* h_1 \to V h_1}^{\lambda_{\gamma^*} \lambda_1 \to \lambda_V \lambda_1}
(s_1,t_1,Q_2^2)  \, .
\label{Two_to_Three}
\end{eqnarray}

The auxiliary amplitude in Eq. (\ref{Two_to_Three}) 
for the emission of a photon
of transverse polarization $\lambda_V$, and transverse momentum
$\bq_1 = - \bp_1$ can be written as:
\begin{eqnarray} 
\bra{p_1', \lambda_1'} J_\mu \ket{p_1, \lambda_1} 
\epsilon_{\mu}^*(q_1,\lambda_V) 
={ (\be^{*(\lambda_V)} \bq_1)  \over \sqrt{1-z_1}} 
\, {2 \over z_1} \, \chi^\dagger_{\lambda'} 
\Big\{  F_1(Q_1^2) 
- { i \kappa_p F_2(Q_1^2) \over 2 m_p } 
( \bsigma_1 \cdot [\bq_1,\bn]) \Big\} \chi_\lambda \, .
\end{eqnarray}
Above $F_1$ and $F_2$ are Dirac and Pauli electromagnetic form factors,
respectively. In Ref.\cite{CSS2015} we have included the Pauli form
factors for a first time.

\section{Selected results for vector quarkonia}

\begin{figure}
\begin{center}
\includegraphics[width=4cm]{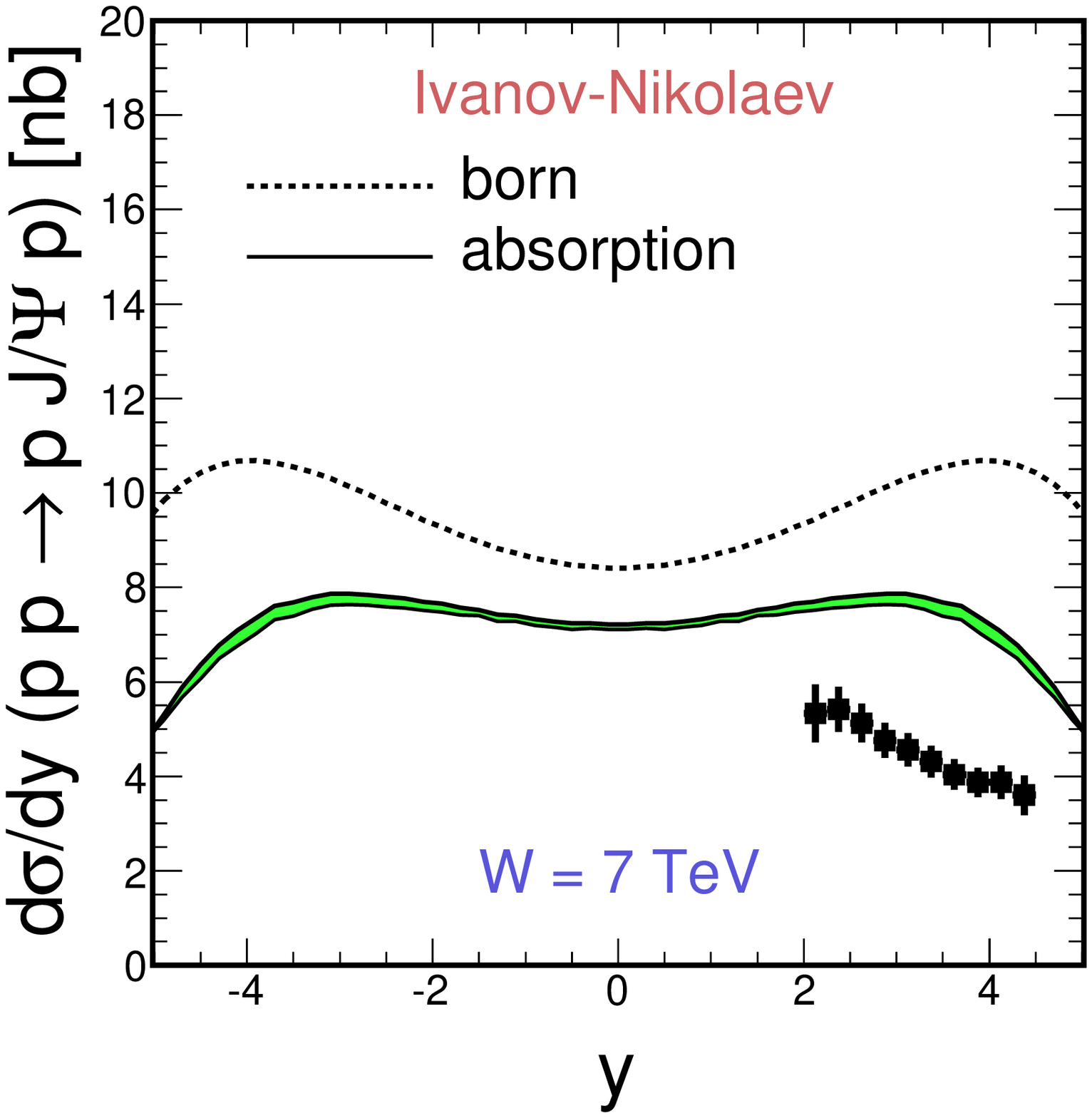}
\includegraphics[width=4cm]{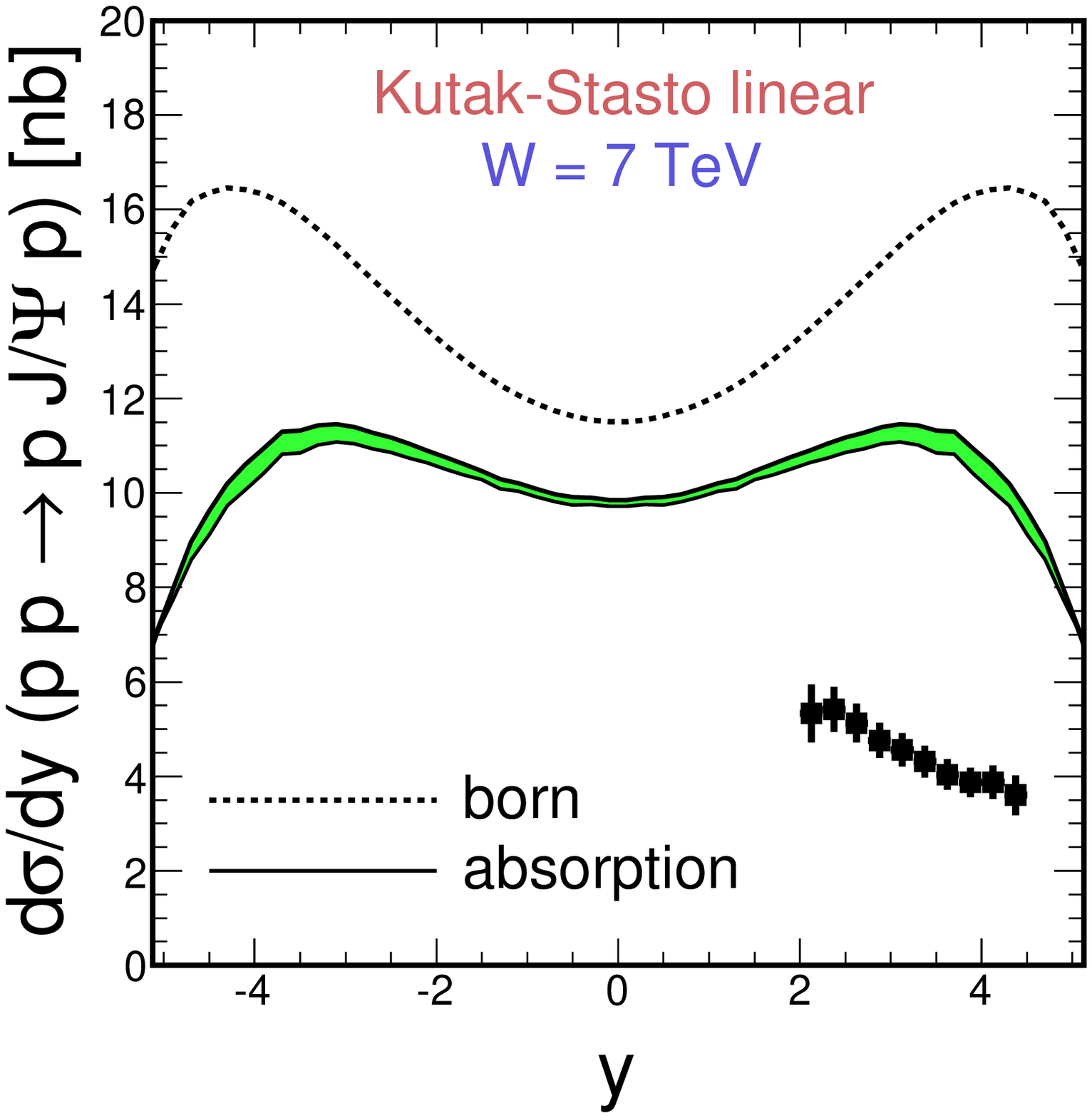}
\includegraphics[width=4cm]{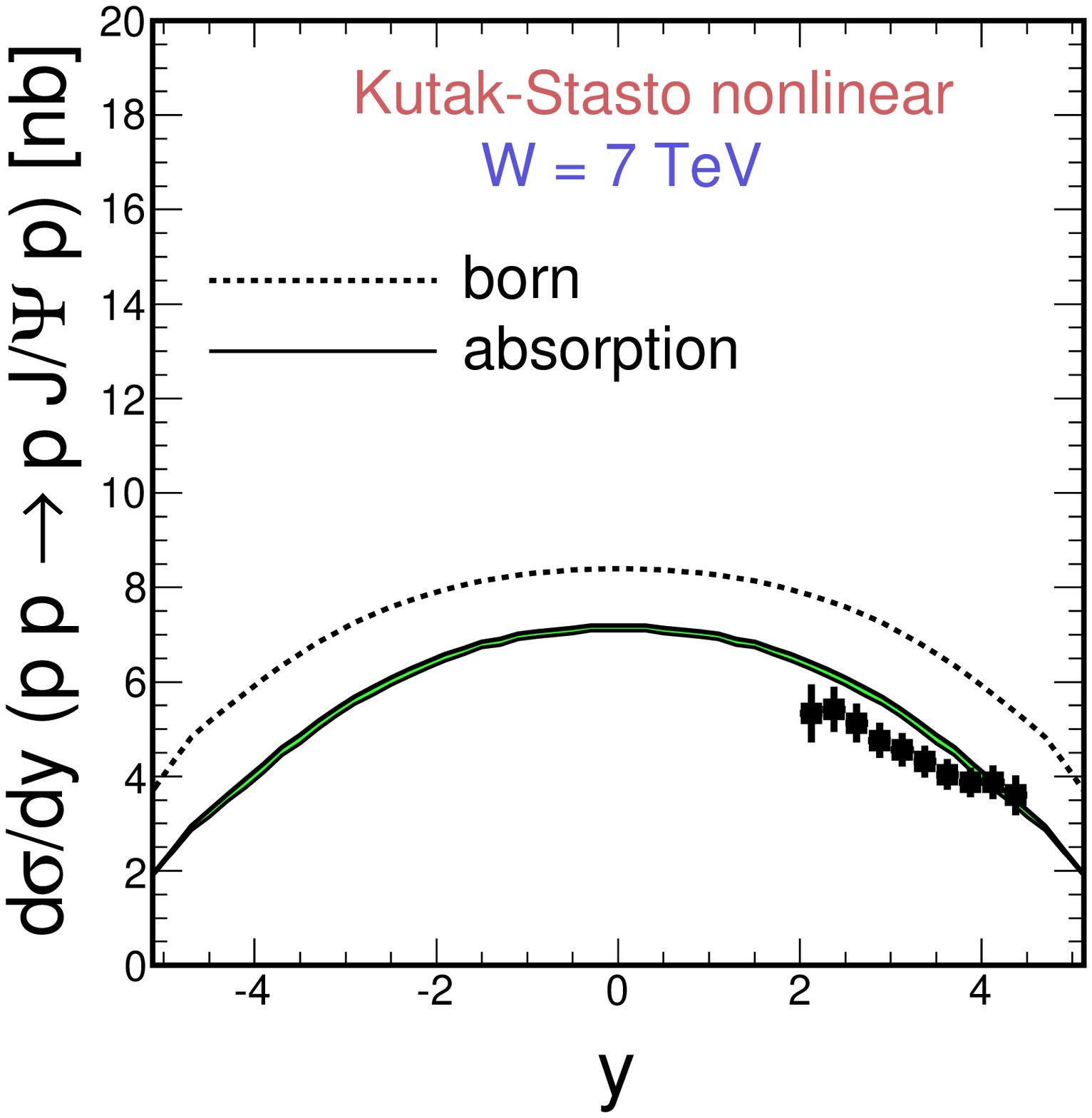}
\end{center}
\caption{Rapidity distributions for three different unintegrated gluon
distributions. The upper curves were obtained in the Born approximation
and the lower band includes absorption effects.}
\label{fig:dsig_dy}
\end{figure}

Rapidity distributions of $J/\psi$ mesons are show in
Fig.\ref{fig:dsig_dy}. In this calculation Gaussian wave functions
were used. Only results with UGDFs that include nonlinear effects
describe the LHCb experimental distributions \cite{LHCb}.
In our opinion it is too preliminary to conclude that we observe 
nonlinear effects or onset of gluon saturation.
In Ref.\cite{CSS2015} we obtained similar distributions for 
$\psi'$ \cite{CSS2015}.

\begin{figure}
\begin{center}
\includegraphics[width=4cm]{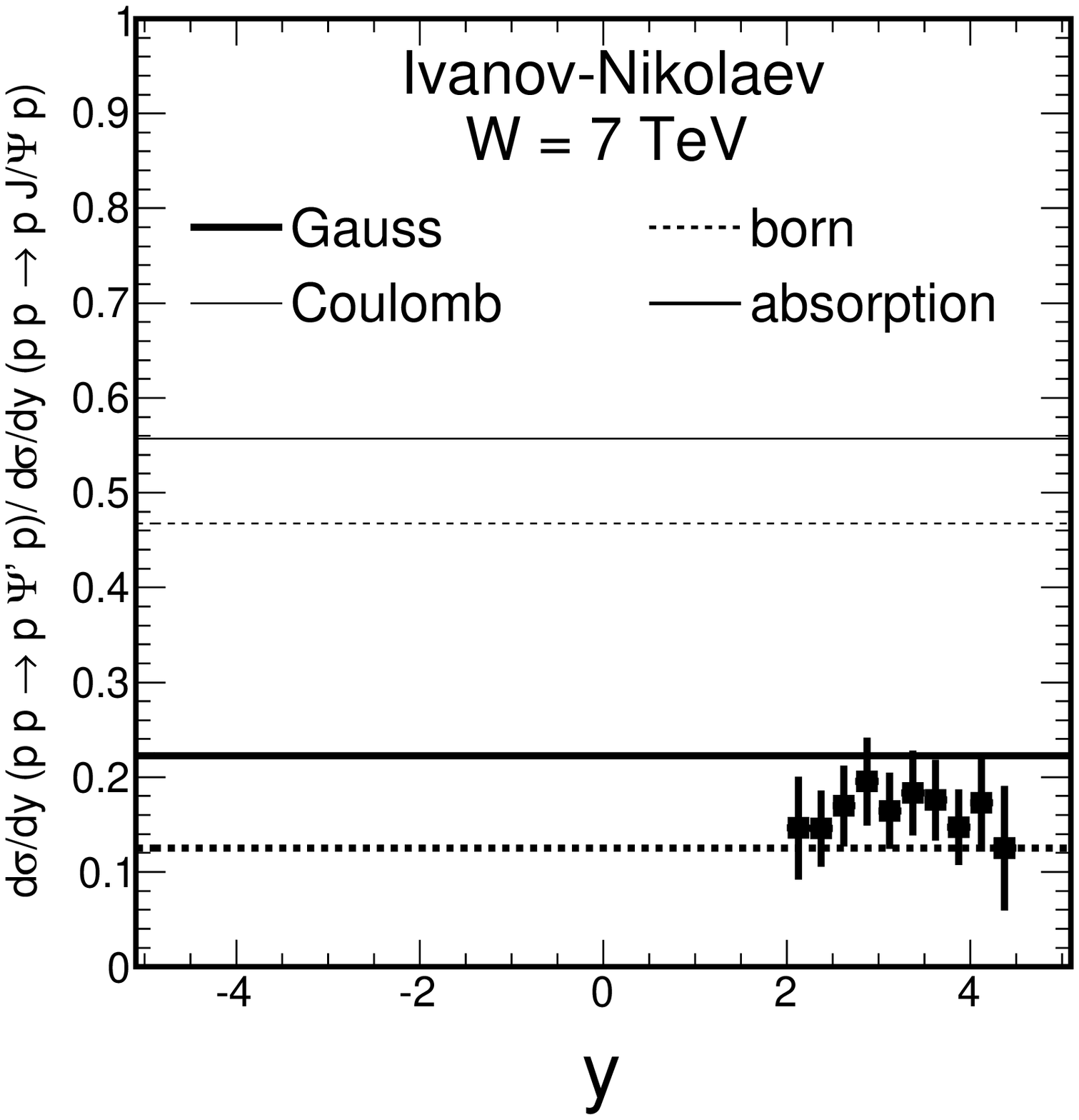}
\includegraphics[width=4cm]{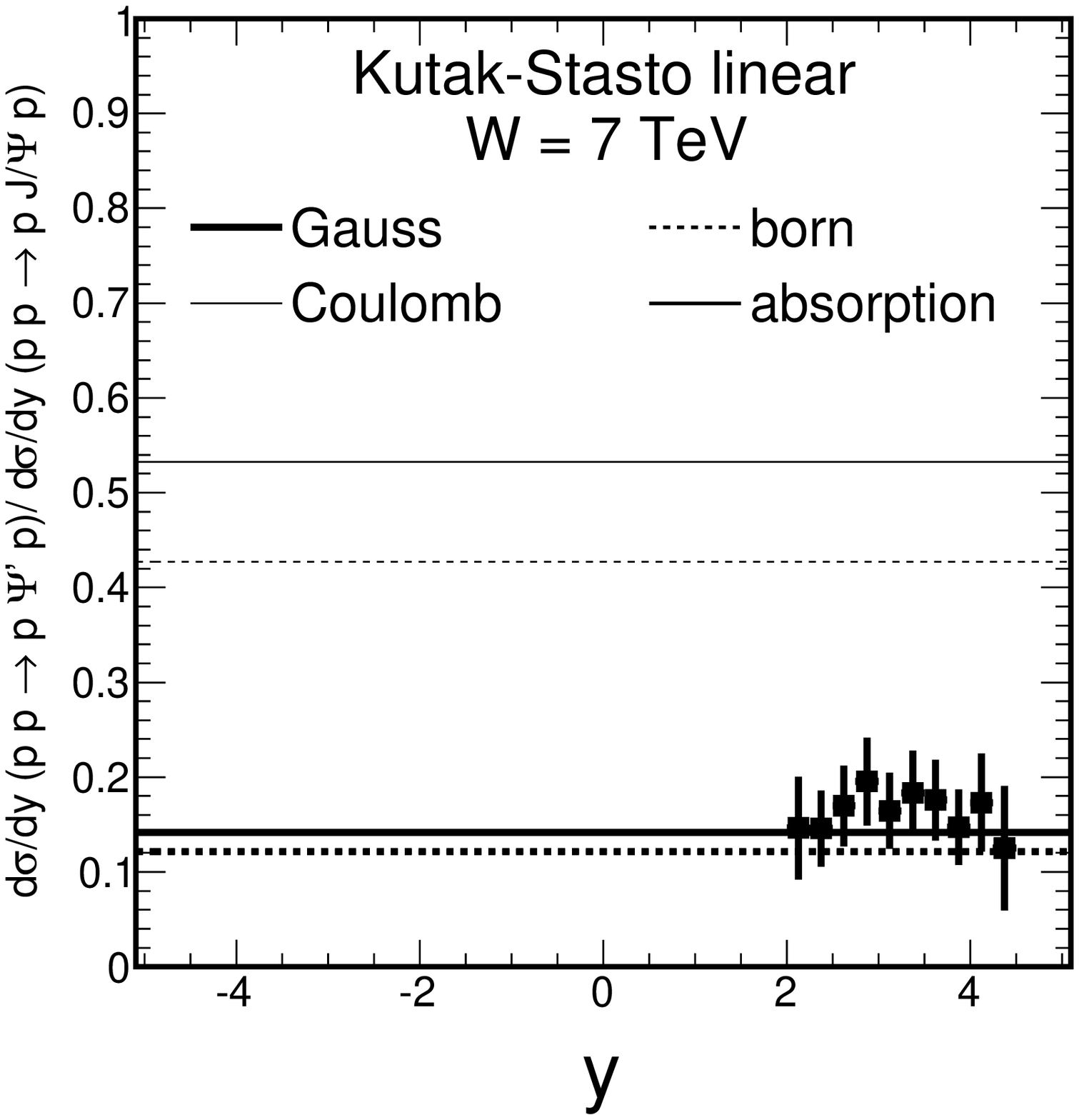}
\includegraphics[width=4cm]{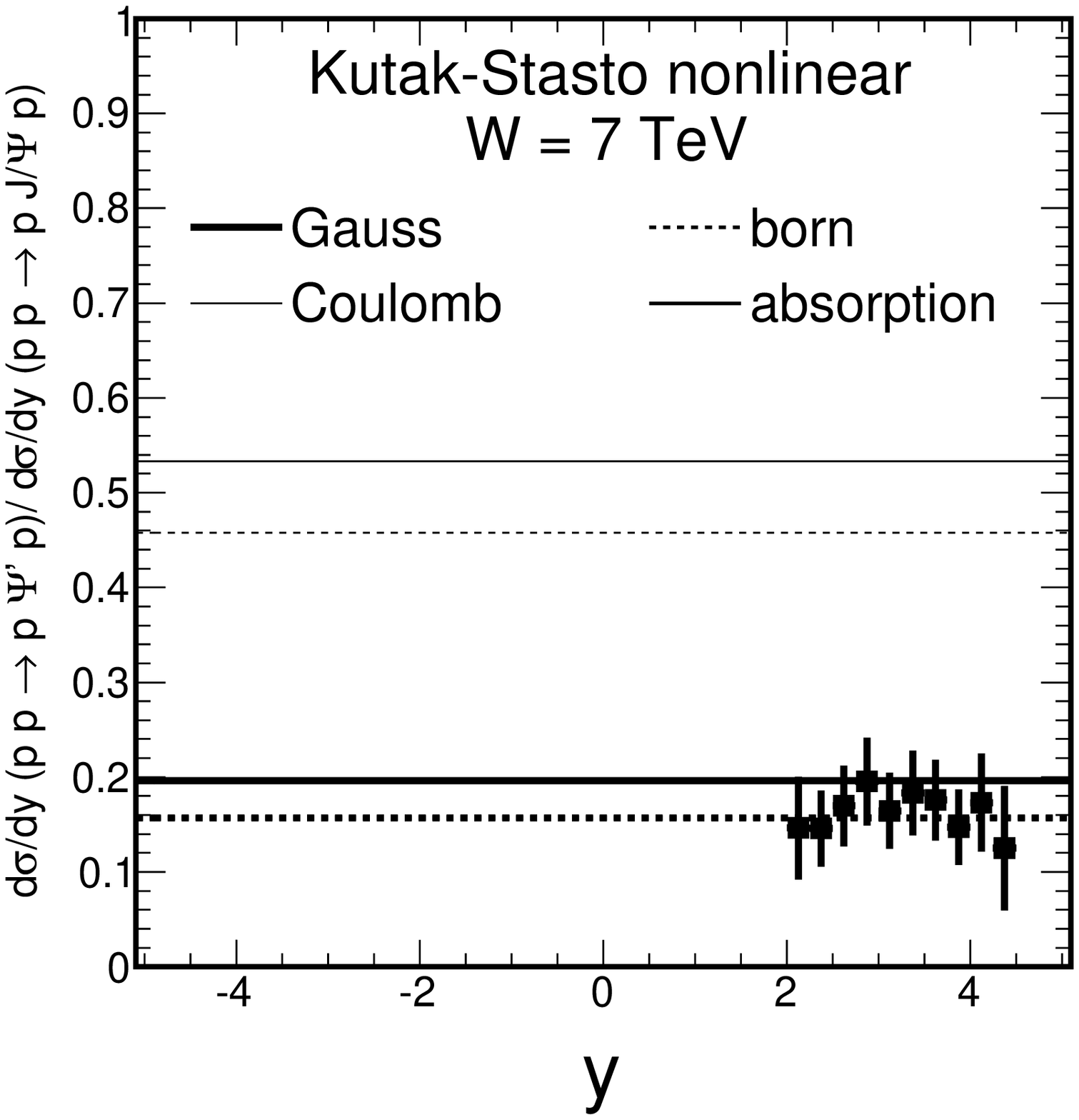}
\caption{The ratio of the cross section for $\psi'$ and $J/\psi$.}
\label{fig:ratios}
\end{center}
\end{figure}

Very interesting quantity is the ratio of the cross sections
for $\psi'$ and $J/\psi$ production. As shown in Fig.\ref{fig:ratios}
such a ratio is very sensitive to the functional form of the $c \bar c$
light-cone wave function.
We conclude that the Gauss $c \bar c$ wave function much better 
describes the LHCb data \cite{LHCb} than Coulomb WF. In some calculation
in the literature the wave functions are not included explicitly and only
point-like coupling is used \cite{Ryskin-Martin}.

\section{$\gamma \gamma$ production of dileptons}

\begin{figure}
\begin{center}
\includegraphics[width=3.5cm]{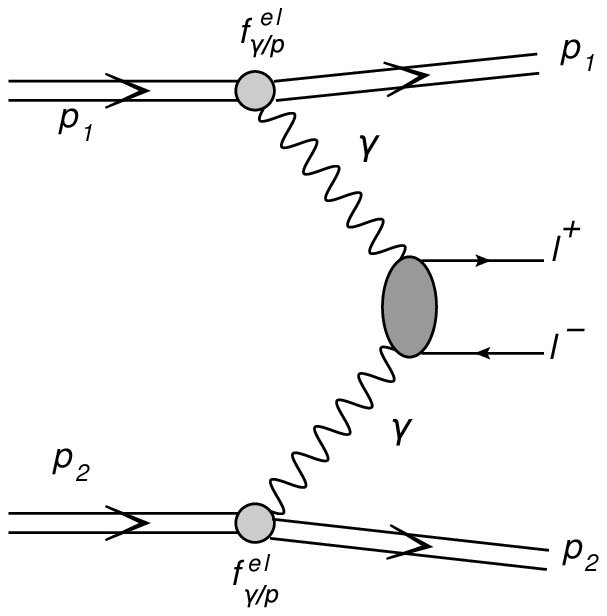}
\includegraphics[width=3.5cm]{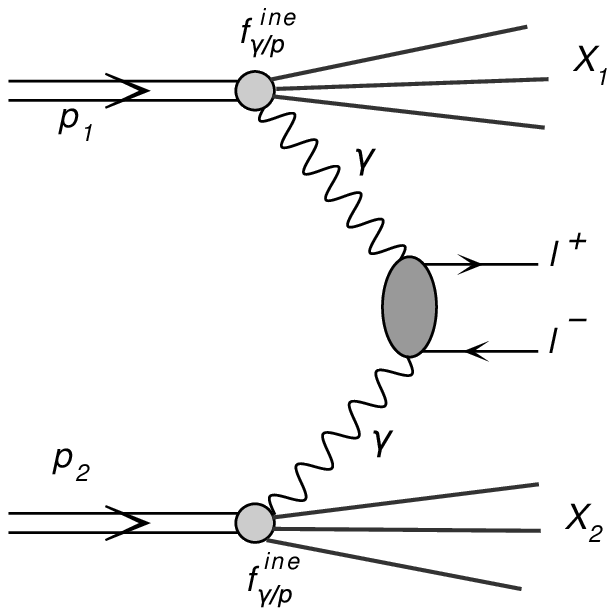} \\
\includegraphics[width=3.5cm]{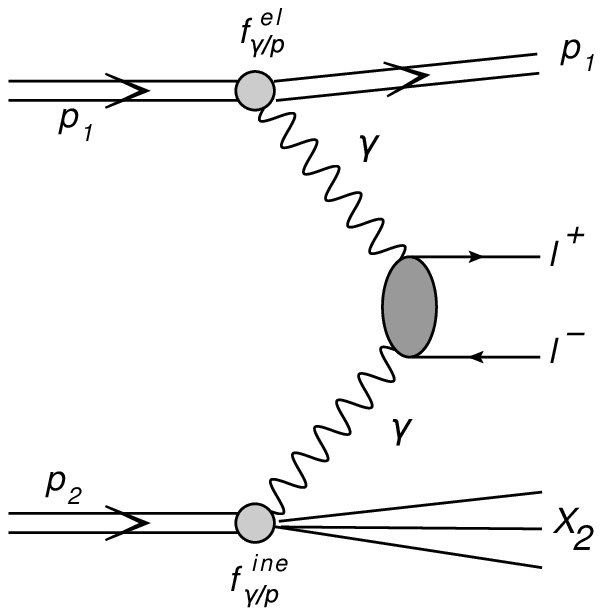}
\includegraphics[width=3.5cm]{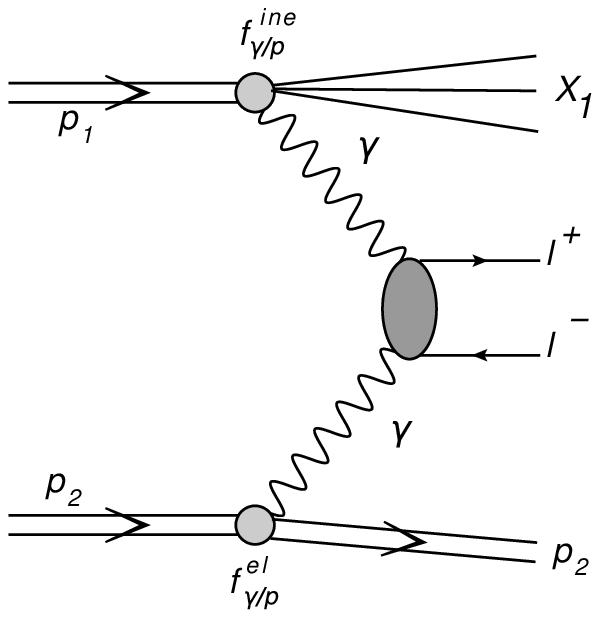}
\end{center}
\caption{The different categories of $\gamma \gamma$ processes.}
\label{fig:diagrams_gammagamma}
\end{figure}

The $\gamma \gamma$ processes can be categorize according
to the final state (see Fig.\ref{fig:diagrams_gammagamma}).
All the processes were discussed in \cite{SFPSS2015,LSS2015}.

Two different approaches were discussed in the context of $\gamma
\gamma$ production of dileptons.

In the collinear approach the corresponding cross sections are calculated 
respectively as:
%
%
%
\begin{eqnarray}
 \frac{d \sigma^{(k,l)}}{d y_1 d y_2 d^2p_t} &=& \frac{1}{16 \pi^2 {\hat s}^2}
  x_1 \gamma_{k}(x_1,\mu^2) \; x_2 \gamma_{l}(x_2,\mu^2) \; 
\overline{|{\cal M}_{\gamma \gamma \to l^+l^-}|^2} \, ,
\end{eqnarray}
where $k,l \in \{\rm{el},\rm{in} \}$ stand for the processes with intact 
($\rm{el}$) or dissociated $\rm{in}$ proton at the photon vertex. 
The $\gamma_k(x,\mu^2)$ are the corresponding photon distributions in a proton,
with or without the condition of breakup.
The elastic photon distributions are calculated with the help of 
the nucleon electromagnetic  form factors, while the inelastic distributions
can be obtained by using a combined QCD/QED evolution \cite{QED-PDF}.

In the $k_t$-factorization, proposed recently in \cite{SFPSS2015,LSS2015},
the fully unintegrated photon flux can be written as (below incoming particles are denoted as $A,B$):
\begin{eqnarray}
{d {\cal{F}}_{\gamma^* \leftarrow A} (z,\bq,M_X^2) \over dM_X^2}  = {\alpha_{\rm{em}}\over \pi}  \, (1-z) \, 
\Big( {\bq^2 \over \bq^2 + z (M_X^2 - m_A^2) + z^2 m_A^2  }\Big)^2  \, \cdot {p_B^\mu p_B^\nu \over s^2} \, W_{\mu \nu}(M_X^2,Q^2)  \, .
\end{eqnarray}
Information on the virtual photon-proton interaction is contained in the hadronic tensor, which is obtained in terms of the electromagnetic currents as:
\begin{eqnarray}
 W_{\mu \nu}(M_X^2,Q^2) = \overline{\sum_X} (2 \pi)^3 \, \delta^{(4)} (p_X - p_A - q) \, \bra{p} J_\mu \ket{X}\bra{X} J_\nu^\dagger \ket{p} \, d\Phi_X \, ,
\label{eq:Wmunu}
\end{eqnarray}
%
The virtual photoabsorption cross sections are related to hadronic tensors as:
\begin{eqnarray}
 \sigma_T(\gamma^* p) = {4 \pi \alpha_{em} \over 4 \sqrt{X}} \, \Big(- {\delta^\perp_{\mu\nu} \over 2} \Big)  2\pi W^{\mu \nu}(M_X^2,Q^2) \, , \, 
 \sigma_L(\gamma^* p) = {4 \pi \alpha_{em} \over 4 \sqrt{X}} \, e^{0}_\mu e^{0}_\nu \, 2 \pi W^{\mu \nu}(M_X^2,Q^2) \, .
\end{eqnarray}
It is customary to introduce dimensionless structure function $F_i(x_{\rm Bj},Q^2), i = T,L$ as
\begin{eqnarray}
 \sigma_{T,L}(\gamma^* p) = {4 \pi^2 \alpha_{em} \over Q^2} \, {1 \over \sqrt{1 + {4 x^2_{\rm Bj} m_A^2 \over Q^2}} } \, F_{T,L}(x_{\rm Bj},Q^2) \, ,
\end{eqnarray}
At high energies, in the calculation of photon fluxes, the contribution of the structure function
\begin{eqnarray}
F_2(x_{\rm Bj},Q^2)  &=& { F_T(x_{\rm Bj},Q^2) +F_L(x_{\rm Bj},Q^2) \over 1 + {4 x^2_{\rm Bj} m_A^2 \over Q^2}} 
\end{eqnarray}
dominates.

The unintegrated fluxes enter the cross section for dilepton production as
\begin{eqnarray}
 {d \sigma^{(i,j)} \over dy_1 dy_2 d^2\bp_1 d^2\bp_2} &&=  
\int  {d^2 \bq_1 \over \pi \bq_1^2} {d^2 \bq_2 \over \pi \bq_2^2}  
{\cal{F}}^{(i)}_{\gamma^*/A}(x_1,\bq_1) \, 
{\cal{F}}^{(j)}_{\gamma^*/B}(x_2,\bq_2) 
{d \sigma^*(p_1,p_2;\bq_1,\bq_2) \over dy_1 dy_2 d^2\bp_1 d^2\bp_2} \, , 
\label{eq:kt-fact}
\end{eqnarray}
where $i,j \in \{ \rm{el},\rm{in} \}$ again refer to elastic and inelastic processes on the proton sides.
For brevity we integrated over invariant masses of the possible dissociated system.

The longitudinal momentum fractions carried by photons can be obtained from rapidities and
transverse momenta of the charged leptons as:
\begin{eqnarray}
x_1 = \sqrt{ {\bp_1^2 + m_l^2 \over s}} e^{y_1} 
    + \sqrt{ {\bp_2^2 + m_l^2 \over s}} e^{y_2} 
\; , 
x_2 = \sqrt{ {\bp_1^2 + m_l^2 \over s}} e^{-y_1} 
    + \sqrt{ {\bp_2^2 + m_l^2 \over s}} e^{-y_2} \, .
\end{eqnarray}
%

\section{Selected results for $l^+ l^-$ production}

\begin{figure}
\begin{center}
\includegraphics[width=5cm]{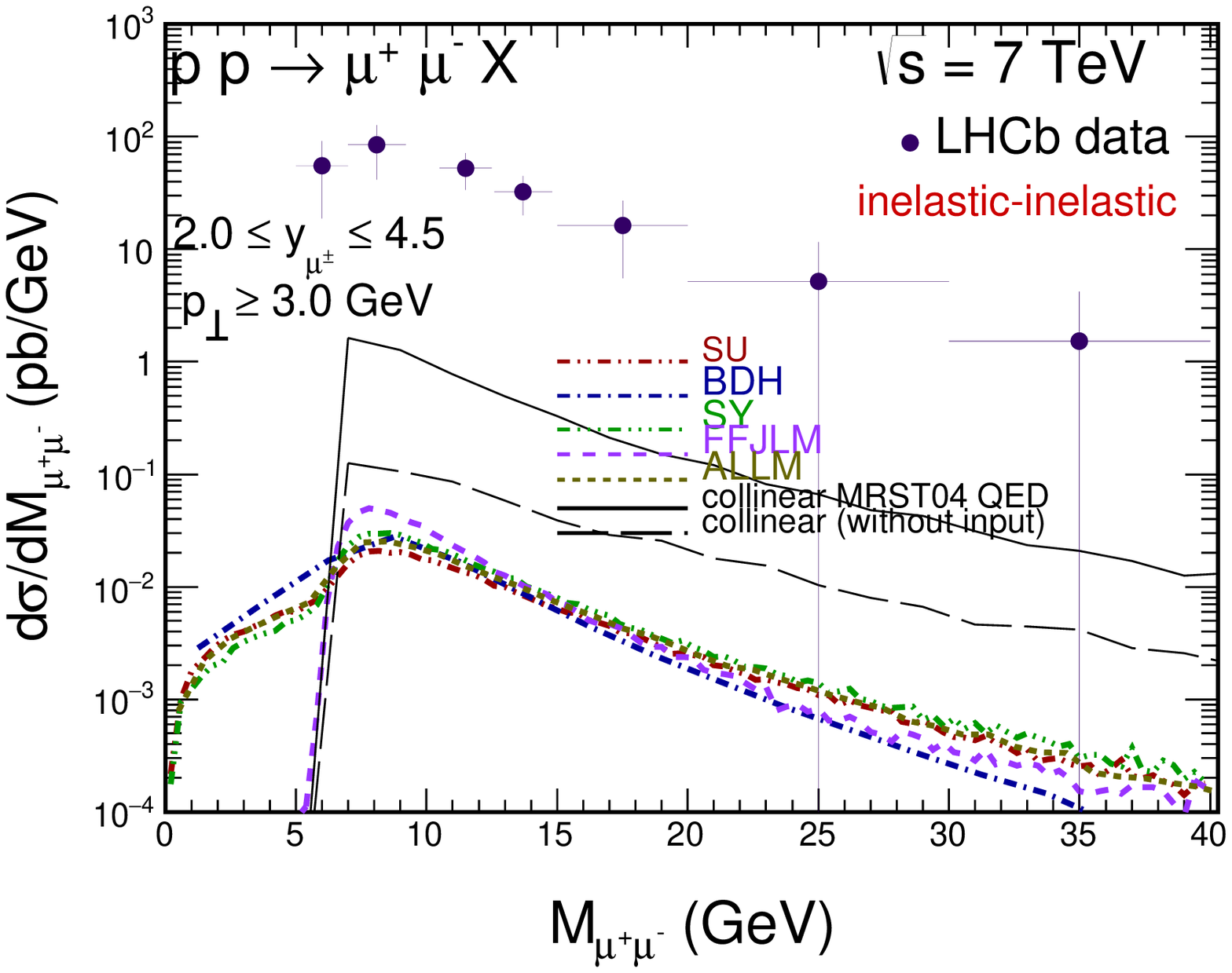}
\includegraphics[width=5cm]{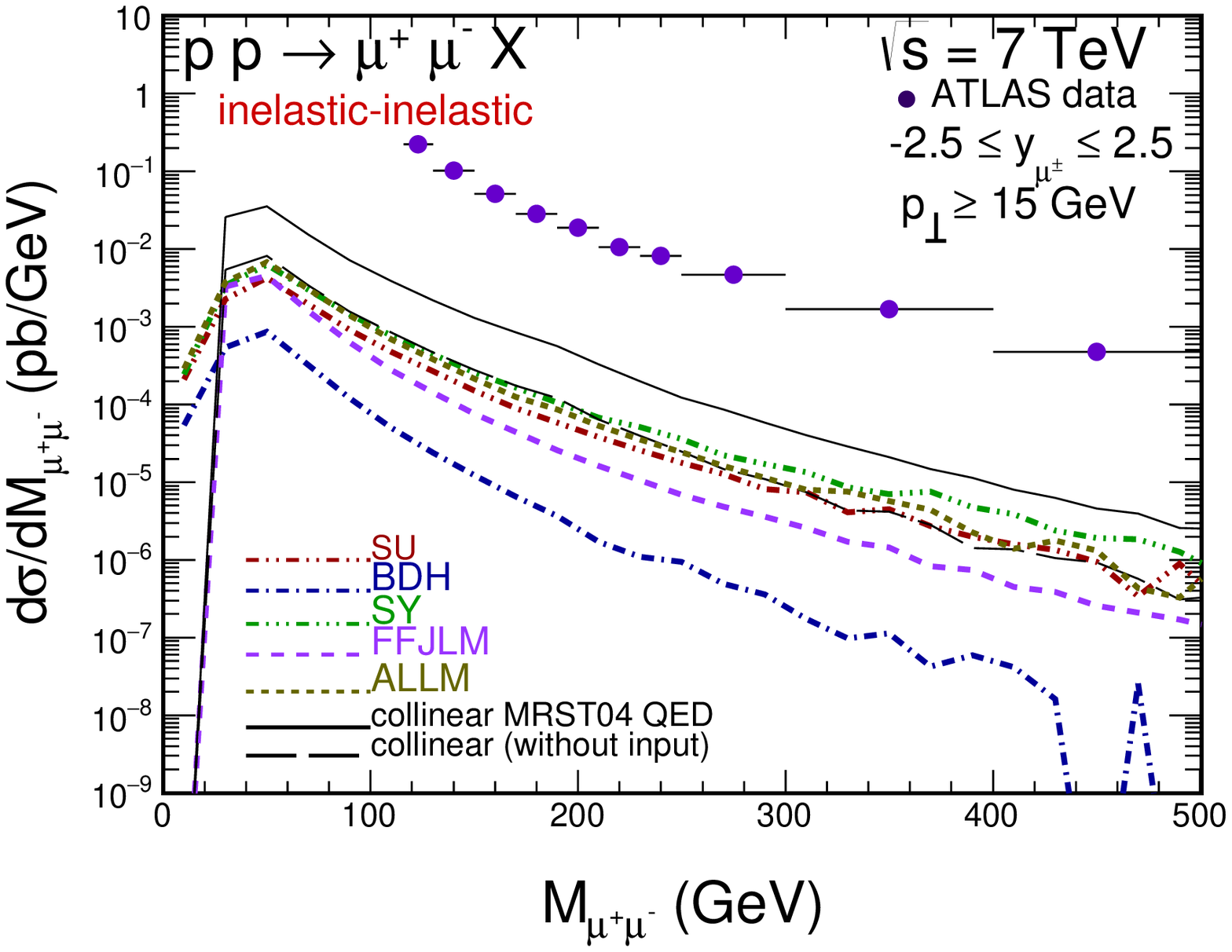} \\
\includegraphics[width=5cm]{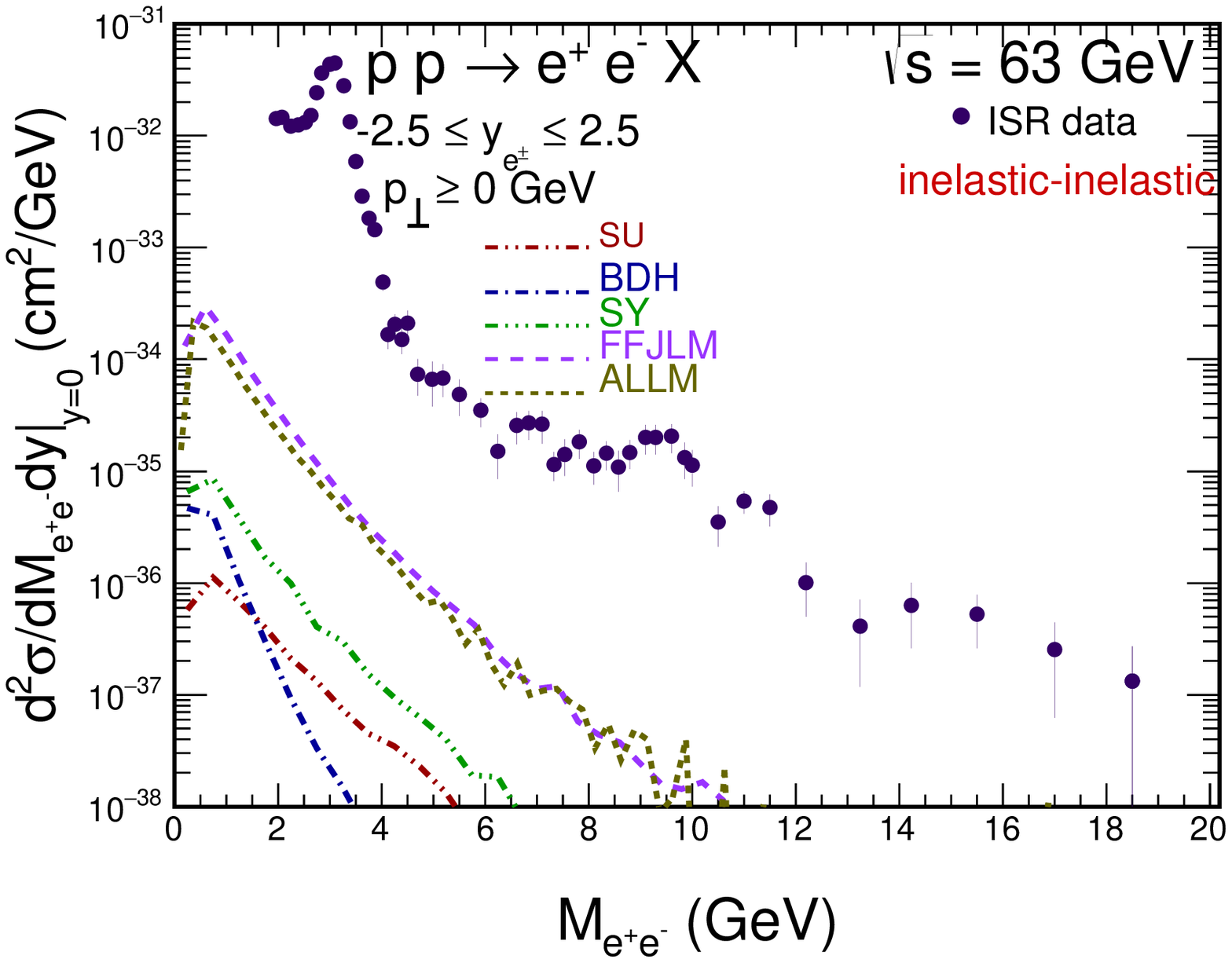}
\includegraphics[width=5cm]{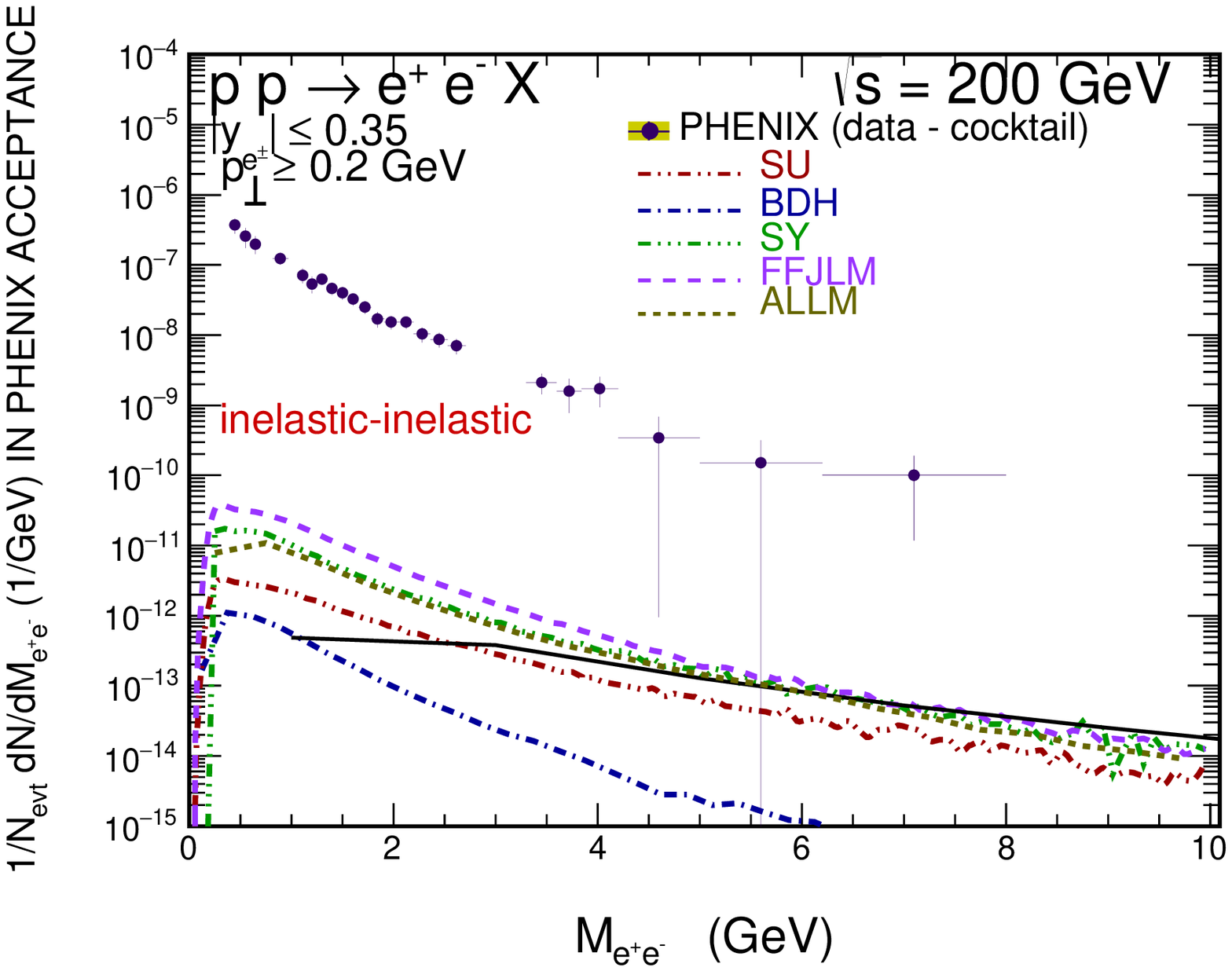}
\end{center}
\caption{Dilepton invariant mass distributions for different
experimental conditions (see the four panels) and for different
structure functions (see different lines).
}
\label{fig:dsig_dMll_inin}
\end{figure}

As an example in Fig.\ref{fig:dsig_dMll_inin} we show invariant mass
distributions for double dissociative processes (both protons undergo
electromagnetic dissociation). The calculations have been performed
for different experimental conditions specified in the figure legend.
We show results for different proton structure functions from the
literature. The different structure functions lead to quite different
results. The presented results strongly depend on the kinematical
regions of longitudinal momentum fraction and photon virtuality
($x$,$Q^2$) which were not sufficiently well studied experimentally
and theoretically in which an interplay of perturbative 
and nonperturbative effects takes place. 
We observe that the $\gamma \gamma$ contribution constitutes
only a small fraction of the cross section compared to the experimental data and 
Drell-Yan contribution (not shown explicitly here). 
Similar results for elastic-inelastic channel were shown in 
\cite{LSS2015}.

\begin{figure}
\begin{center}
\includegraphics[width=5cm]{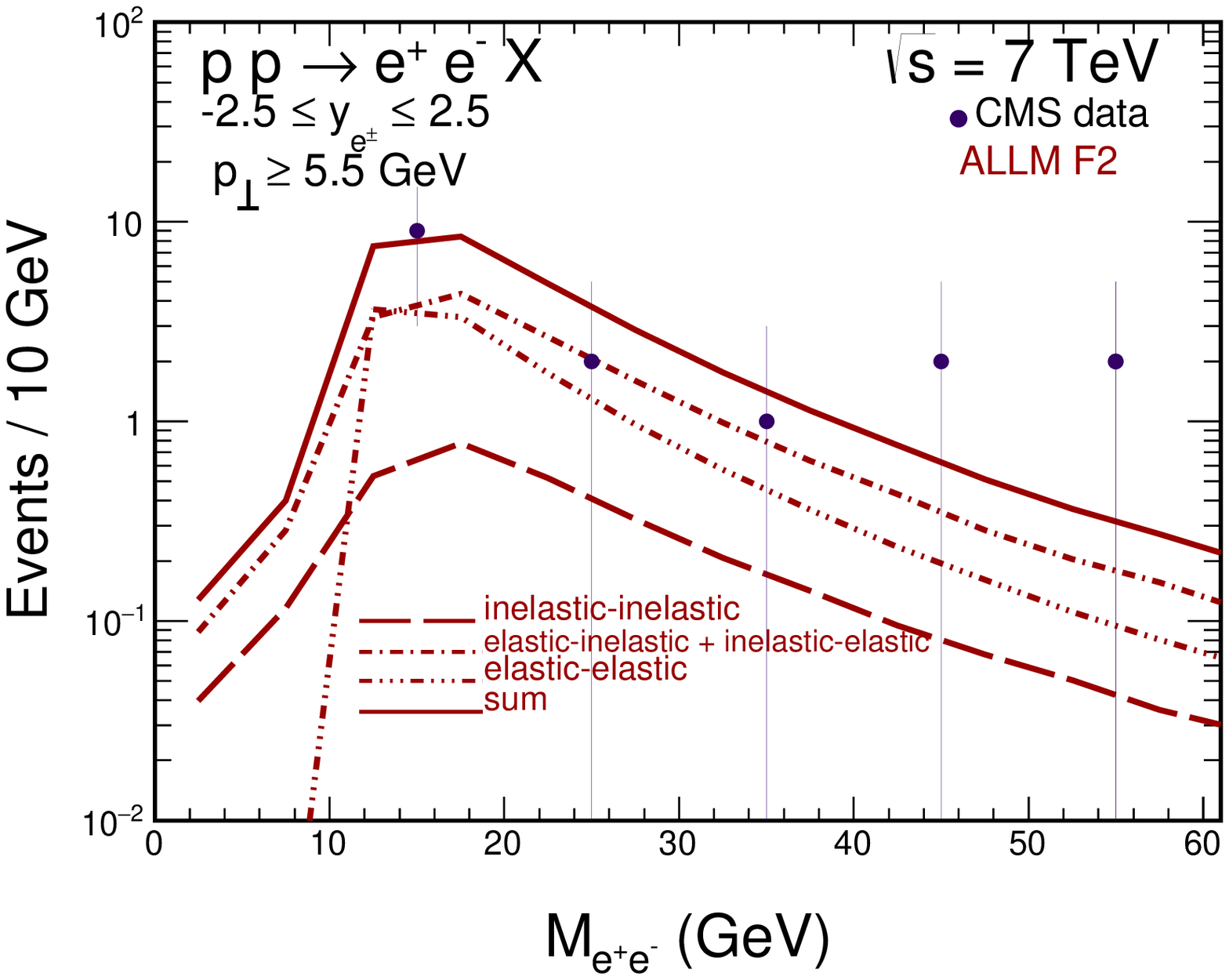}
\includegraphics[width=5cm]{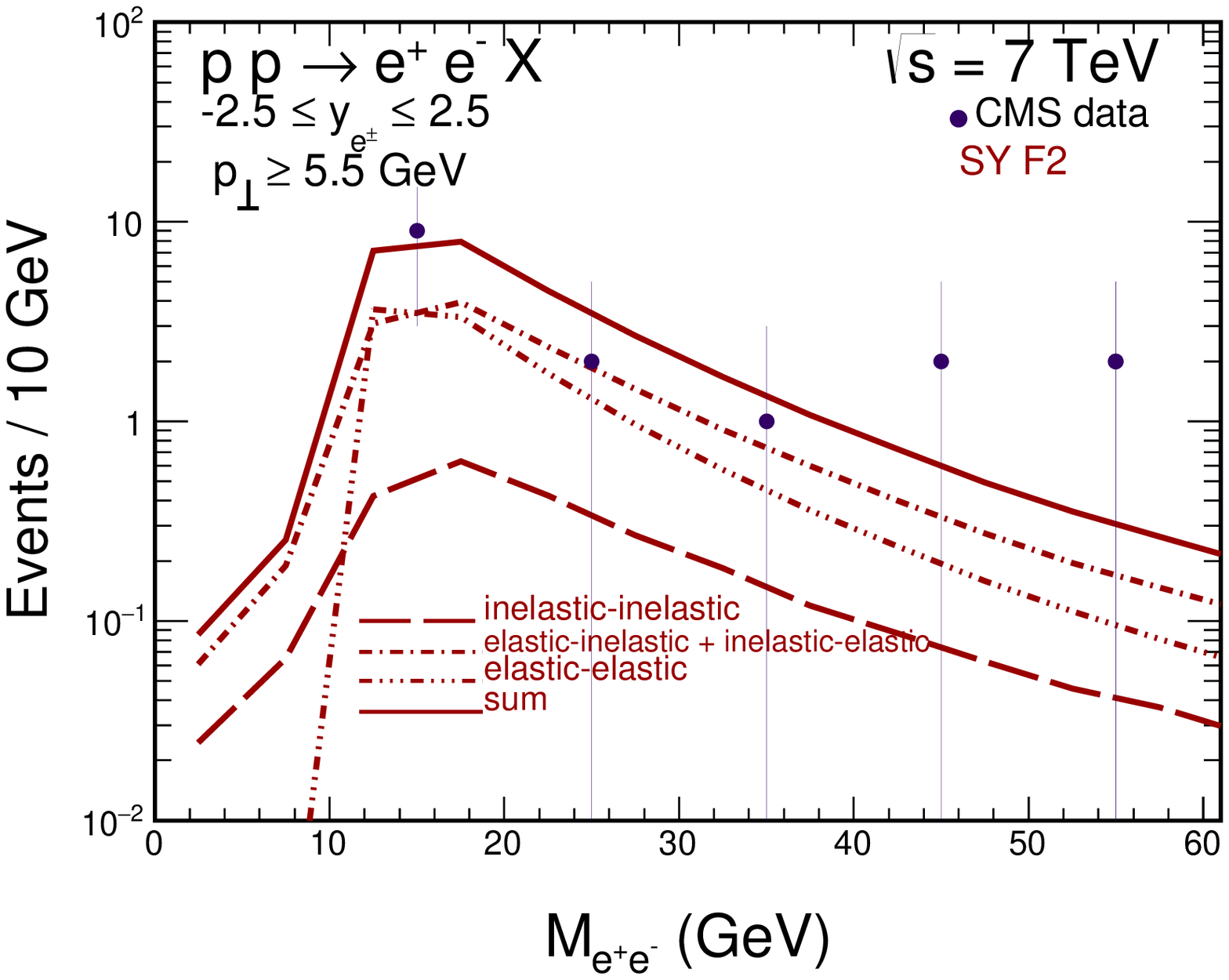}
\end{center}
\caption{Dilepton invariant mass distributions for two (best) 
structure functions. The CMS data points are shown for comparison.}
\label{fig:CMS_dsig_dMll}
\end{figure}

\begin{figure}
\begin{center}
\includegraphics[width=5cm]{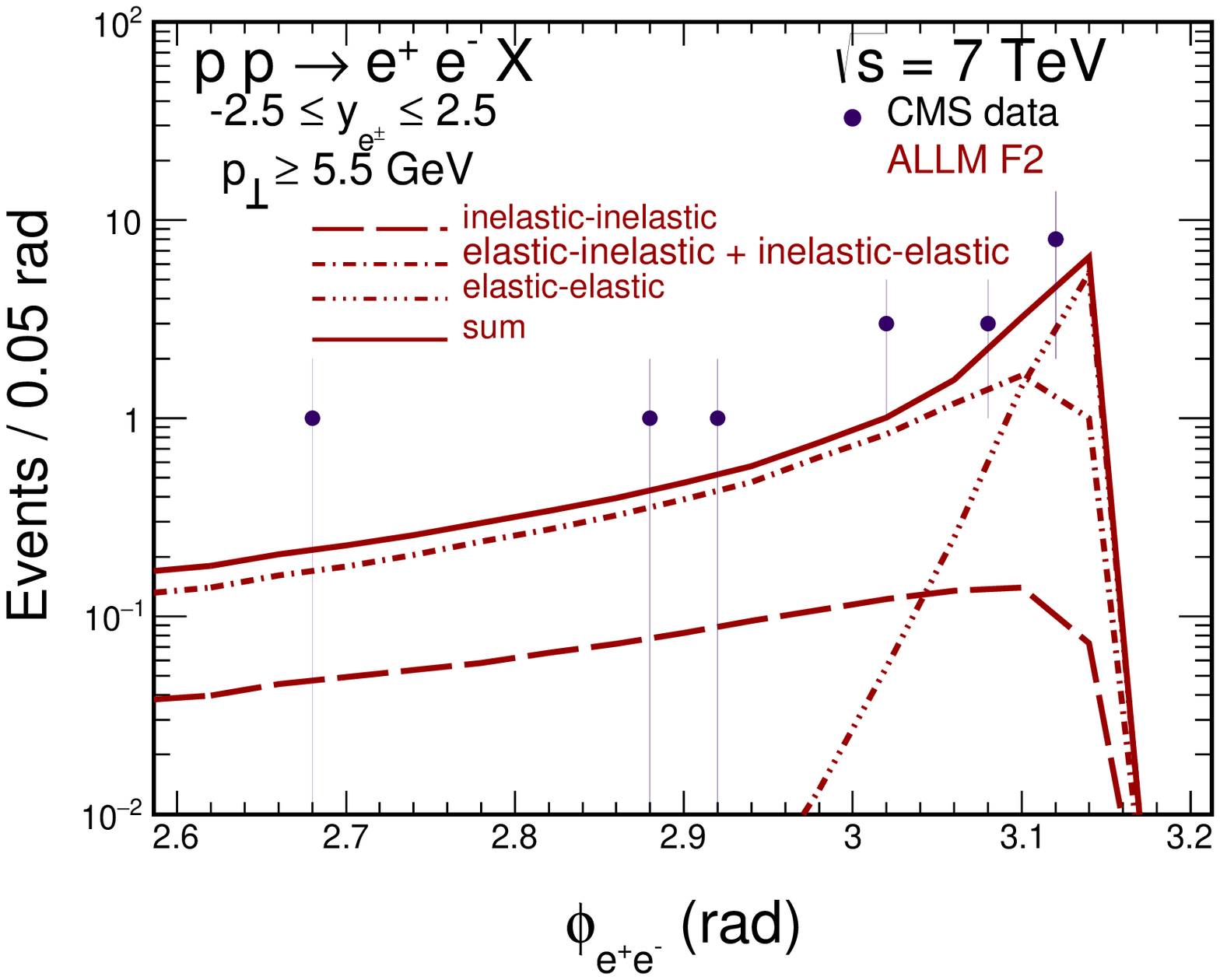}
\includegraphics[width=5cm]{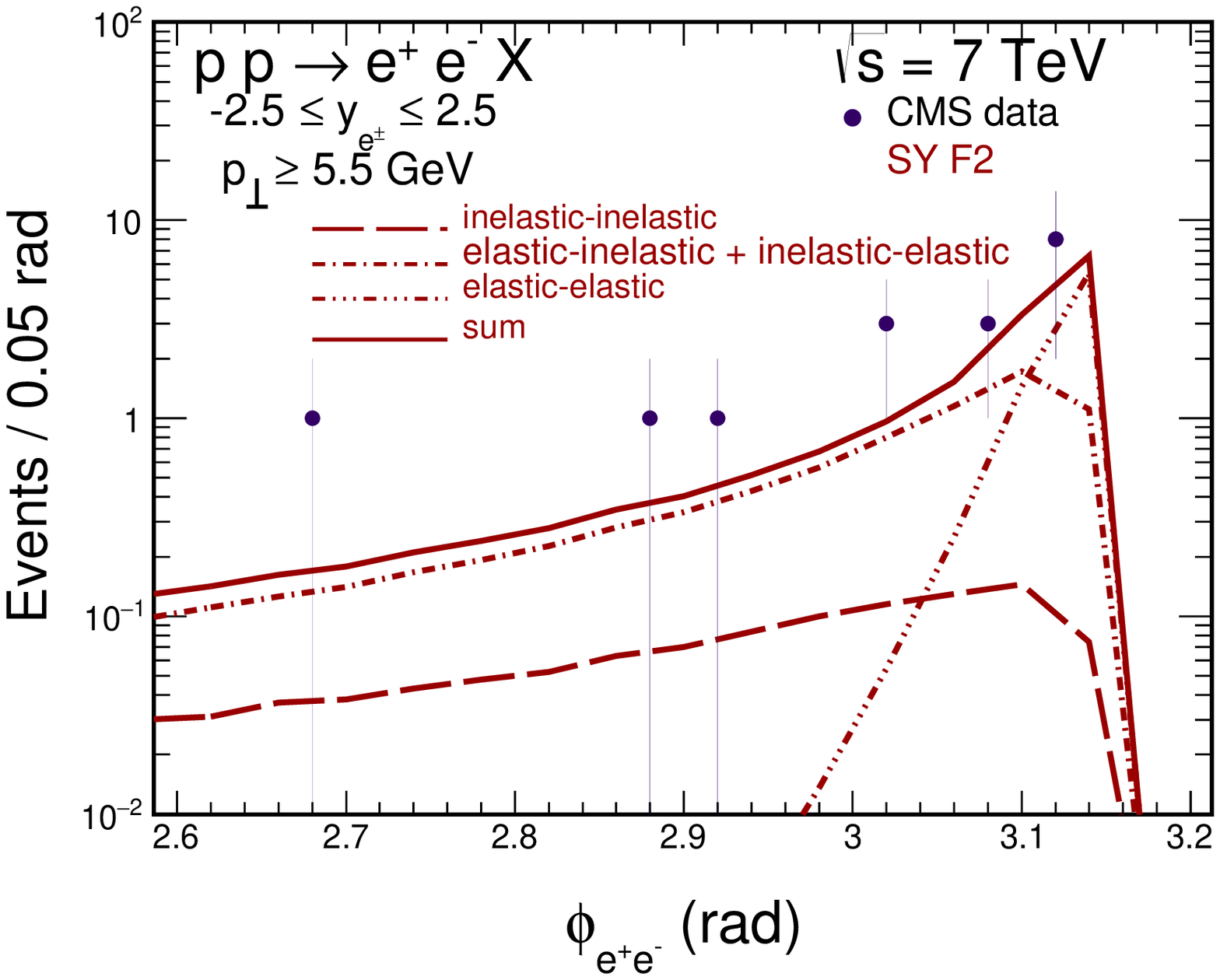}
\end{center}
\caption{Distribution in relative azimuthal angle
between leptons for two (best) structure functions.
The CMS data points are shown for comparison.}
\label{fig:CMS_dsig_dphi}
\end{figure}

\begin{figure}
\begin{center}
\includegraphics[width=5cm]{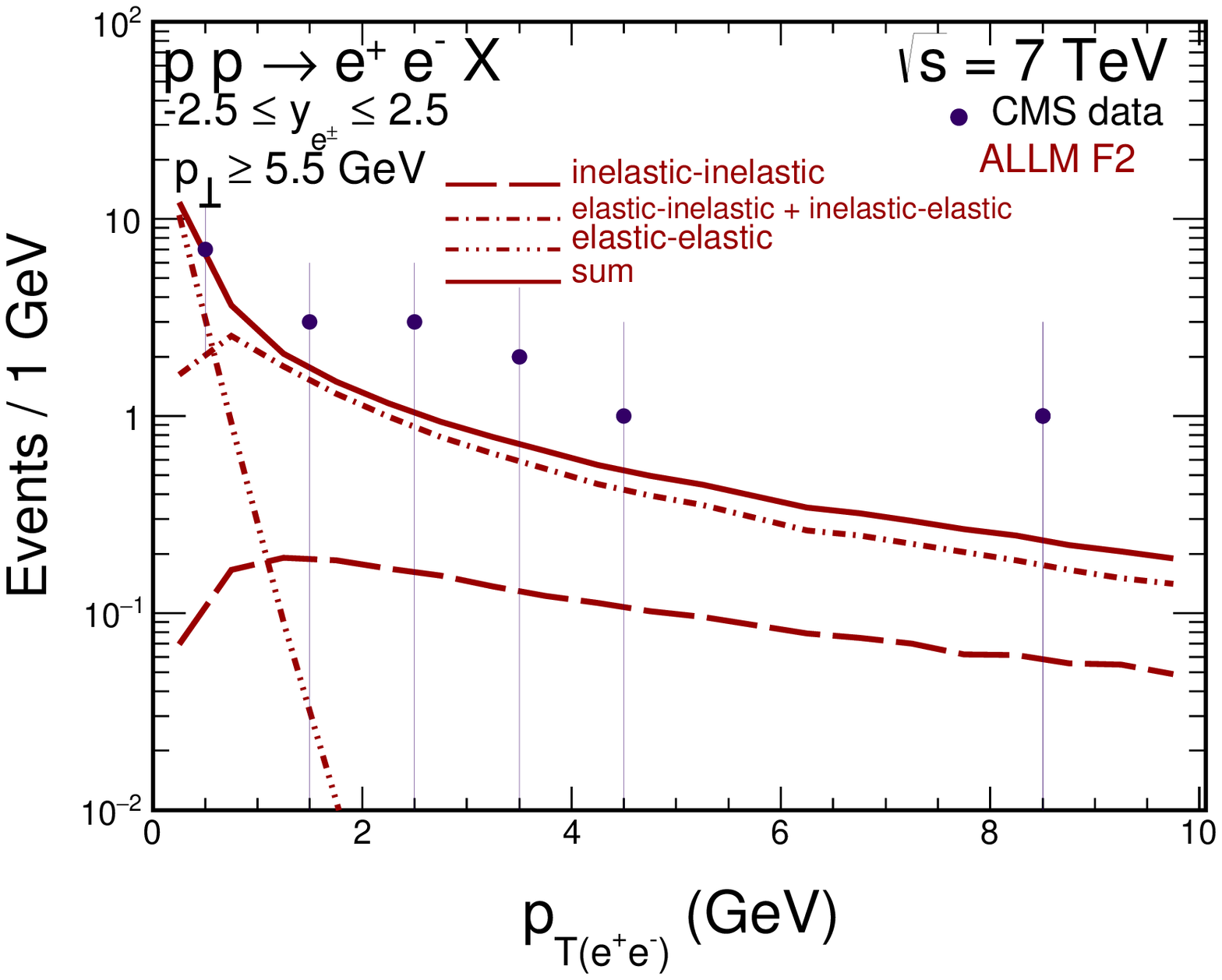}
\includegraphics[width=5cm]{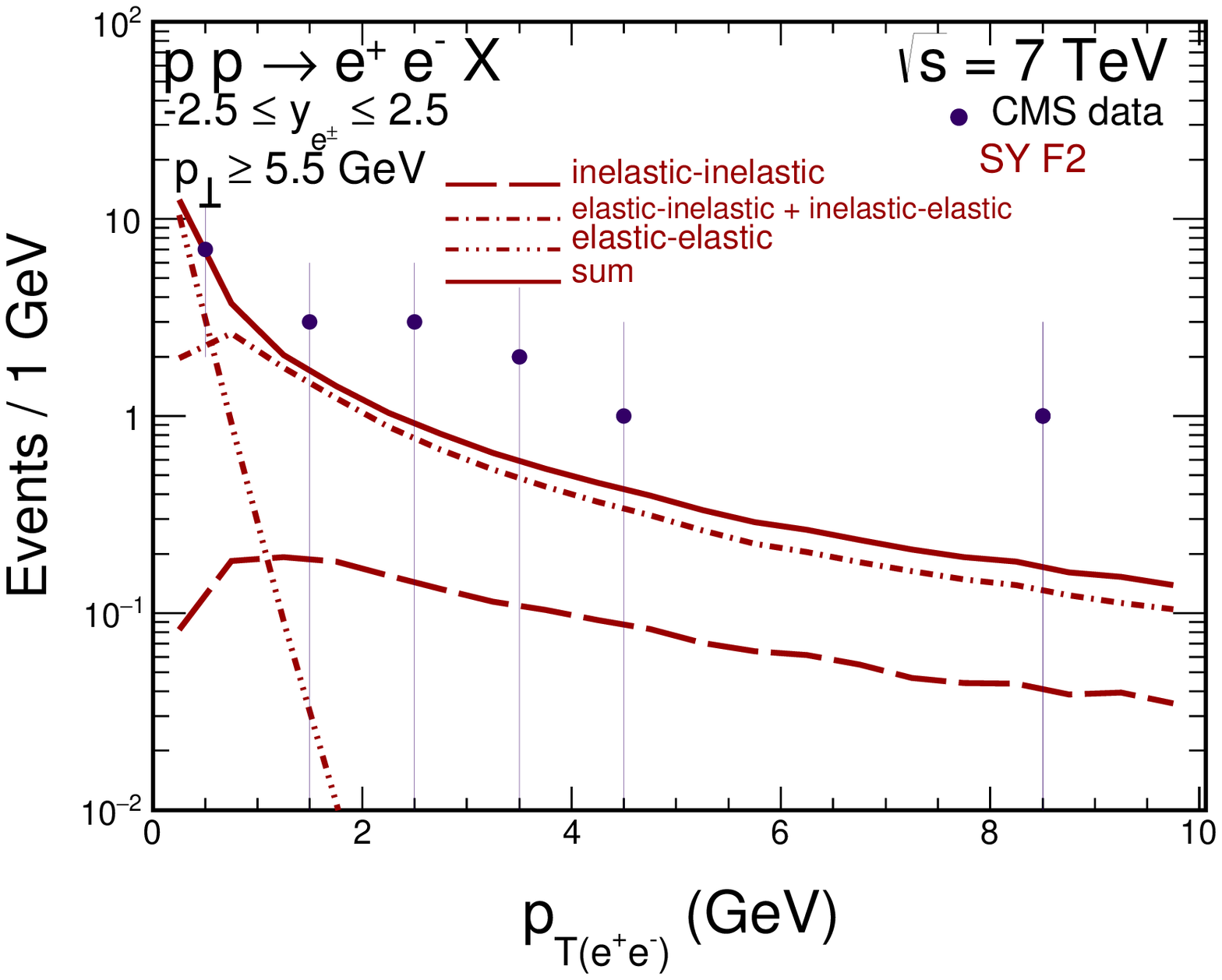}
\end{center}
\caption{Distribution in transverse momentum of the pair of leptons
for two (best) structure functions.
The CMS data points are shown for comparison.}
\label{fig:CMS_dsig_dptpair}
\end{figure}

Now we wish to show results which can be directly compared to
experimental data \cite{CMS}. The data were obtained by imposing
a kinematical constraint on lepton isolation.
The results are shown in Figs.\ref{fig:CMS_dsig_dMll},\ref{fig:CMS_dsig_dphi},
\ref{fig:CMS_dsig_dptpair}. A relatively good agreement has been
achieved. A better quality results are expected from Run II at the LHC.

\section{Conclusions}

Our results and open problems for charmonia production 
can be summarized as follows.
We have found some model dependent indication of presence of 
nonlinear effects in the small-$x$ gluon distribution of a proton.
It is important to remember, that the present experiments are not fully exclusive 
and rather semi-inclusive processes have to be studied.
Experimentally so far proton dissociation has been "extracted" in a model 
dependent way assuming some functional form in $p_t$.
From HERA we have some limited knowledge only about diffractive dissociation.
Compared to exclusive production of $J/\psi$ at HERA,
in $pp$ collisions there is also electromagnetic dissociation.
Interference effects due to the two diagrams in Fig. \ref{fig:diagrams_exclusive_Jpsi} were predicted.
It would be nice to see a modulation in $\phi_{pp}$ due to
interference effects between the two diagrams. 
In the future, the CMS+TOTEM and ATLAS+ALFA experiments
could measure purely exclusive reaction and study dependence on many more variables.

Our results for photon-photon induced processes can be summarized as:
We discussed two different approaches for $\gamma \gamma$ processes --
collinear and $k_t$-factorization, they are not completely equivalent.
We have found strong dependence on the structure function input
to the photon fluxes in the $k_t$-factorization approach.
Semi-exclusive contributions with dissociation are large
which is an interesting lesson for $p p \to p pJ/\psi$.
The photon-photon contributions are, however, rather small compared 
to the Drell-Yan contribution.
A reasonable description of the CMS data with isolated electrons was obtained 
(recently also the ATLAS collaboration obtained similar results).
So far only the collinear approach has been applied e.g. to the
$p p \to (\gamma \gamma) \to W^+ W^- X Y$ processes.
Such a reaction is important in searches for Beyond Standard Model
effects. A large cross section was found \cite{LSR2015}. 
A calculation in $k_t$-factorization approach would be very valuable.



\end{document}